# Data management to support reproducible research

B. A. Wandell
A. Rokem
L. M. Perry
G. Schaefer
R. F. Dougherty

Center for Cognitive and Neurobiological Imaging
Stanford University
Stanford, CA 94305

Correspondence: Wandell@stanford.edu

**Acknowledgements**:  We gratefully acknowledge support from the Simons Foundation, the Weston-Havens Foundation, and NSF-BCS 1228397.  We thank Ryan Chamberlain, Joyce Farrell, and J. Gomez for comments on the manuscript.

## Abstract

We describe the current state and future plans for a set of tools for scientific data management (SDM) designed to support scientific transparency and reproducible research. SDM has been in active use at our MRI Center for more than two years.  We designed the system to be used from the beginning of a research project, which contrasts with conventional end-state databases that accept data as a project concludes.   A number of benefits accrue from using scientific data management tools early and throughout the project, including data integrity as well as reuse of the data and of computational methods.

## Introduction

Reproducible research is the idea that the product of scientific research is not only the paper, but also the data and software needed to reproduce the results (Buckheit & Donoho, 1995; Fomel & Claerbout, 2009). Modern data management tools such as searchable databases and computational notebooks are valuable tools for scientific project management. Here we describe a set of tools for scientific data management (SDM, previously called NIMS) that are designed to support reproducible research.

SDM is currently used at Stanford's Center for Neurobiological Imaging (CNI), an MRI Center that serves more than 400 users in 40 labs.  The SDM archiving and database tools have been in active use since 2012, and they continue to evolve. We are adding significant functionality for computational method reuse and sharing in the next phase of the project.

It is useful to compare the differences between SDM and other neuroimaging databases. The conventional database is designed to serve as a stable, centralized system that aggregates a large collection of measurements. Data are typically contributed to the system at the end of a research study, when the analyses are done and as the research is published (Poline et al., 2012, Figure 1). Such end-state databases are frequently organized to support a particular research topic. Three successful centralized systems are the Alzheimer's Disease Neuroimaging Initiative (ADNI, (Jack et al., 2008)), the National Database for Autism Research (NDAR, (Hall, Huerta, McAuliffe, & Farber, 2012; NIH, 2015)), and the Human Connectome Project (HCP, (Van Essen et al., 2013)).

In contrast, we designed SDM to be used from the beginning and throughout the project. At the CNI, data from every scan is immediately entered into SDM at which time it is archived and processed. A number of benefits accrue from using database tools early in the scientific process. For example, SDM gathers the raw data along with metadata from the headers, placing them in a database with a searchable and clearly defined organization. Also, the database and associated files are backed-up, and thus the user's raw data are secure from the outset of the project. In this way, SDM ensures data integrity and eliminates the possibility of data loss.

A second difference is that SDM is engineered to support research broadly rather than a specific research topic. This design increases the likelihood that investigators working in different fields might share best practices in data management and common computational practices.

Given our goal of using SDM from the start of the scientific process, it was logical to build in capabilities for performing certain basic data operations: lossless data compression, conversion from DICOM to NIfTI format, and simple data visualizations. Also, data entered into SDM are automatically analyzed to assess instrument and data quality. These quality assurance (QA) processes are run in the background and do not require any user intervention. The QA reports are stored alongside the data and are summarized for the MRI Center staff as well as for the users. These methods are applied when data are first acquired, sparing individual labs from creating their own implementations.

In conventional databases, placing data into the centralized system implies that the user is sharing the data widely. This is inappropriate for the first stage of a project, and the SDM design does not require data sharing. In the initial phase of a project, only the PI's group has access to the data. To promote reproducible research and to simplify the task of sharing, and reuse of the data, SDM includes an extensive set of user-rights management tools. The sharing tools are in a simple browser-based interface. The PI sets the data access rules, and data can be shared with an individual, a group, or can be made publicly available to the broader scientific community. Thus, we meet the SDM design goal of reproducible research by implementing an extensive set of simple browser-based tools for user-rights management.

SDM includes data from many different investigators with a broad array of research objectives. SDM uses the database search facility to find potentially useful data that can increase the power or extend the analyses of any one particular study.  It is possible to search for a type of data (e.g., diffusion, anatomical, functional) or to find subjects with certain characteristics (e.g., age, gender) or subject metadata (for example pertaining to a specific research topic, such as reading development or mood disorders).  An investigator can discover that data matching the search query exists within SDM even if access is not currently granted.  The investigator can then contact colleagues and ask about data reuse.  If an agreement is reached, the browser-based user-rights tools can easily provide specific access to the data.

The primary content in SDM comprises data archived directly from the MRI scanner at the CNI. These data are automatically collected, formatted, and stored in SDM without any user interaction. There are cases, however, when we wish to add data from other sources to a SDM database.  This is done either by placing raw files manually into a directory that is continuously monitored by SDM or by uploading the data using the SDM browser-based interface. In this way, older datasets, or datasets collected at other institutions, have been incorporated.

The SDM database and functions are structured to be capable of automatically collecting, formatting, and storing data from a range of scientific instruments without requiring user interaction. The initial application is for MRI, but the software architecture is designed to be extensible to other instruments and data types with well-defined file formats.

## Implementation

SDM consists of several key software components.

*API:* The SDM API serves as the program center and the gateway to the SDM database. The API has been engineered to adhere to the tenets of the REST (REpresentational State Transfer) framework and architectural style (Masse, 2011).

*Database:* The SDM database uses MongoDB - a NoSQL, cross-platform, document-oriented database. MongoDB is flexible because it uses dynamic schemas (Chodorow, 2013).

*Reaper:* Reaper is the point of contact between the Instrument and SDM. Reaper is engineered in Python and directly communicates with a scientific instrument (e.g., an MRI scanner), from which it 'reaps' every data file that is captured by the instrument (e.g., DICOM files and raw k-space data) for sorting and processing in the SDM database. Reaper communicates with the rest of the SDM pipeline via the SDM API. Importantly, Reaper allows data to enter SDM without user interaction.  The raw data are captured and sorted into the database by the Reaper, eliminating concerns about user compliance.  The Reaper shifts the burden of capture and archiving from the user to the software.

*Processor:* Once raw data from the instrument are reaped they are sorted into the database and the SDM data processor, a custom-engineered, open-source library, applies certain processing steps. At present, the library has tools to convert raw DICOM data from GE or Siemens scanners to a widely used, compressed, file format - NIfTI. Quality assurance metrics are part of the library, and the results of these analyses are included in the database with the raw data. There is a visualization module built on PanoJS (Migurski & Allen, 2005) to allow users to view data within the SDM UI. The SDM data processor is modular, so that researchers can design and contribute their own purpose-built computational modules that plug-in to SDM data processor and perform specialized processing operations.

*User Interface:* Data management and user-rights are managed through a Web UI engineered using HTML5 and AngularJS. The SDM UI allows users to access their data from anywhere in the world and provides the interface through which they can browse, organize, download, search and share their data. Users may also search for and request access to data at remote sites. Users can create Virtual Projects, consisting of data across projects, labs, and even institutions. Sharing and permissions are managed within the SDM UI. Access to the SDM UI is managed via OAuth.

*Federation:* The SDM software includes peer-discovery, cross-site database queries, and data sharing. A SDM instance can broadcast its existence to a web site that securely provides across-site access to resources.

## Current state and plans for SDM

The current SDM release operating at the CNI manages more than 250 million DICOM files, comprising 71,000 scans (e.g., a diffusion measurement with multiple directions and b-values, T1-weighted anatomical, or a functional series). There are about 9000 scans for subjects between 10-18 years of age, 35,000 scans for subjects from 19-30, 15000 from 31-50, and 8500 from 51-90. Amongst these, there are 3500 diffusion data sets; 29,000 functional MRI scans; 6300 T1-weighted anatomical data, 380 MR-spectroscopy measurements, 470 perfusion measurements, and a wide assortment of measurements obtained as part of sequence development protocols. Whenever a new MR pulse sequence is introduced, the data (e.g., DICOM, P-files, Physio) will be classified by its metadata and then placed in the database. If recon code exists for this class, the data will be transformed using this code.

The next SDM release, planned for 2015-Q2, introduces several new features. One important new feature flows from the search mechanism. Having identified a set of data, and further having obtained access permission, the user can collect the data into a 'virtual experiment'. The data in a collection can be treated as if they were acquired as part of a real experiment, in that user-rights management tools and data analysis tools can be applied equally to original data or to a collection.

A second new feature supports sharing between sites: Users will be able to search data and manage user-rights across separate SDM instances.

New computational features will be incorporated using a format that supports user-supplied computations (https://www.docker.com). Investigators will be able to create and apply their computational methods to SDM data, and the results can be stored in the SDM database. These methods will be tracked as part of the scientific data management process, so that users can reuse the original data as well as the complete set of processing steps. Making the data and computations available is a key goal of reproducible research.

## Related data management systems

Several other neuroimaging data management systems have overlapping functionality with SDM. In the following summary, we try to explain the distinctive design goals and features of these several systems. All modern software strives for several key features: an intuitive interface, secure data storage, and the ability to scale. A principal objective of the tools listed here is to help with the centralization and standardization of data.

### XNAT

The Extensible Neuroimaging Archive Toolkit (XNAT) is designed to support both daily lab use and large-scale data archiving, including both single and multi-site archives. The overlap and difference with SDM is clarified by considering the expected workflow. For a typical user, files are transferred from the MR scanner into a pre-archive where they are validated and perhaps processed. After the user is satisfied, the data are place into XNAT's secure central archive. Distinguishing user-owned local copies from archive copies is a fundamental strategy used by XNAT to maintain data integrity (Marcus, 2014; Marcus, Olsen, Ramaratnam, & Buckner, 2007, Fig. 4).

### LORIS

The LORIS informatics systems is architected to automate the flow of clinical trials and complex multi-center studies. To support this functionality, LORIS is written as a subject-centric project with the ability to carefully check the validity of multiple forms of subject data, including behavioral forms and visualization of MR data. LORIS includes search capabilities that permit investigators to download and analyze data locally, as well as a mechanism to interface LORIS data with external data processing pipelines (Das, Zijdenbos, Harlap, Vins, & Evans, 2011).

### LONI IDA

The Laboratory of Neuro Imaging (LONI) Image and Data Archiving system (LONI IDA) is designed to archive, query, visualize and reuse neuroimaging clinical and neurocognitive data. The IDA stores a large amount of data from multiple sites, including the Alzheimer's Disease Neuro Imaging data set and data from the Michael J. Fox Foundation (Van Horn & Toga, 2009). The principal design goal of the LONI IDA is to provide end-state data by securely pooling and de-identifying data from multiple institutions, while providing the ability to share data among qualified investigators.

COINS

The Mind Research Network developed a COllaborative Informatics and Neuroimaging Suite (COINS), which is a suite of web-based open-source tools to manage studies, subjects, imaging, clinical data, and other assessments (Scott et al., 2011). The system includes a set of user-rights management tools that enable sharing specific datasets between investigators within the Mind Research Network. There is also a set of tools to help users acquire and organize progress during clinical studies.  The authors identify the standardization of the measurements and methods as a strength, and they point to the challenge of customization of data types and computation as a weakness.  The COINS design is significantly influenced by the need to assist users obtaining data for standardized clinical trials.  This is an important application that contrasts with the need in scientific projects for flexibility to incorporate novel measurements and develop computational methods de novo.

# Conclusions

SDM is an open-source project (https://github.com/scitran).  Anyone can download and install the database architecture within their lab or Center.  They can then become part of the federated group. MR Centers can install the software so that it is integrated with data acquisition for all users.

The Simons Foundation and the Weston-Havens Foundation supported SDM development.  We are committed to keeping the essential features of the database that touch the data completely open-source.  We are not in a position to help all laboratories that wish to use SDM, say for help with the installation, updates, or creation of advanced features. To solve this problem, we are investigating developing a commercial venture to offer system installation and support to sustain and spread the adoption of the principles embedded in the design without compromising the development of the open source project.

# References

Buckheit, J. B., & Donoho, D. L. (1995). WaveLab and Reproducible Research *Stanford University, Department of Statistics.* . Stanford, California: Stanford University, Department of Statistics.

Chodorow, K. (2013). *MongoDB: The Definitive Guide, 2nd Edition* (2nd ed.): O'Reilly Media.

Das, S., Zijdenbos, A. P., Harlap, J., Vins, D., & Evans, A. C. (2011). LORIS: a web-based data management system for multi-center studies. *Front Neuroinform, 5*, 37. doi: 10.3389/fninf.2011.00037

Fomel, S., & Claerbout, J. F. (2009). Guest Editors' Introduction: Reproducible Research. *Computing in Science and Engg., 11*(1), 5-7. doi: 10.1109/mcse.2009.14

Hall, D., Huerta, M. F., McAuliffe, M. J., & Farber, G. K. (2012). Sharing heterogeneous data: the national database for autism research. *Neuroinformatics, 10*(4), 331-339. doi: 10.1007/s12021-012-9151-4

Jack, C. R., Bernstein, M. A., Fox, N. C., Thompson, P., Alexander, G., Harvey, D., . . . Weiner, M. W. (2008). The Alzheimer's disease neuroimaging initiative (ADNI): MRI methods. *Journal of Magnetic Resonance Imaging, 27*(4), 685-691. doi: 10.1002/jmri.21049

Marcus, D. S. (2014). N euroimaging Informatics: Tools to Manage and Share Neuroimaging and Related Data. In M. S. Gazzaniga & G. R. Mangun (Eds.), *The Cognitive Neurosciences - Fifth Edition*: MIT Press.

Marcus, D. S., Olsen, T. R., Ramaratnam, M., & Buckner, R. L. (2007). The Extensible Neuroimaging Archive Toolkit: an informatics platform for managing, exploring, and sharing neuroimaging data. *Neuroinformatics, 5*(1), 11-34.

Masse, M. (2011). *REST API Design Rulebook*: O'Reilly Media.

Migurski, M., & Allen, D. (2005). PanoJS3. Retrieved from http://www.dimin.net/software/panojs/

NIH. (2015). National Database for Autism Research.   http://ndar.nih.gov/

Poline, J. B., Breeze, J. L., Ghosh, S., Gorgolewski, K., Halchenko, Y. O., Hanke, M., . . . Kennedy, D. N. (2012). Data sharing in neuroimaging research. *Front Neuroinform, 6*, 9. doi: 10.3389/fninf.2012.00009

Scott, A., Courtney, W., Wood, D., de la Garza, R., Lane, S., King, M., . . . Calhoun, V. D. (2011). COINS: An Innovative Informatics and Neuroimaging Tool Suite Built for Large Heterogeneous Datasets. *Front Neuroinform, 5*, 33. doi: 10.3389/fninf.2011.00033

Van Essen, D. C., Smith, S. M., Barch, D. M., Behrens, T. E., Yacoub, E., Ugurbil, K., & Consortium, W. U.-M. H. (2013). The WU-Minn Human Connectome Project: an overview. *Neuroimage, 80*, 62-79. doi: 10.1016/j.neuroimage.2013.05.041

Van Horn, J. D., & Toga, A. W. (2009). Is it time to re-prioritize neuroimaging databases and digital repositories? *Neuroimage, 47*(4), 1720-1734. doi: 10.1016/j.neuroimage.2009.03.086